\documentclass[sigconf]{acmart}

\setcopyright{none}
\renewcommand\footnotetextcopyrightpermission[1]{}
\acmConference[RobustRecSys @ RecSys '24]{Workshop on Design, Evaluation and Deployment of Robust Recommender Systems at Eighteenths ACM Conference on Recommender Systems}{October 14--18, 2024}{Bari, Italy}

\usepackage{mathtools}
\usepackage{bm}
\usepackage{subcaption}

\DeclareMathOperator{\PR}{\mathbb{P}}

\def\One{\text{\usefont{U}{bbold}{m}{n}1}}
\newcommand*{\one}[1]{\ensuremath{\text{\One}_{#1}}}

\DeclareMathOperator{\mpipe}{\mid}

\begin{document}

\title{Reducing Popularity Influence by Addressing Position Bias}

\author{Andrii Dzhoha}
\email{andrew.dzhoha@zalando.de}
\affiliation{%
    \institution{Zalando SE}
    \city{Berlin}
    \country{Germany}
}

\author{Alexey Kurennoy}
\email{alexey.kurennoy@zalando.ie}
\affiliation{%
    \institution{Zalando SE}
    \city{Berlin}
    \country{Germany}
}

\author{Vladimir Vlasov}
\email{vladimir.vlasov@zalando.de}
\affiliation{%
    \institution{Zalando SE}
    \city{Berlin}
    \country{Germany}
}

\author{Marjan Celikik}
\email{marjan.celikik@zalando.de}
\affiliation{%
    \institution{Zalando SE}
    \city{Berlin}
    \country{Germany}
}

\renewcommand{\shortauthors}{Dzhoha et al.}

\begin{abstract}
    Position bias poses a persistent challenge in recommender systems,
    with much of the existing research focusing on refining ranking relevance and driving user engagement.
    However, in practical applications, the mitigation of position bias does not always result
    in detectable short-term improvements in ranking relevance.
    This paper provides an alternative, practically useful view of what position bias reduction methods can achieve.
    It demonstrates that position debiasing can spread visibility and interactions more evenly across the assortment,
    effectively reducing a skew in the popularity of items induced by the position bias through a feedback loop.
    We offer an explanation of how position bias affects item popularity.
    This includes an illustrative model of the item popularity histogram and the effect of the position bias on its skewness.
    Through offline and online experiments on our large-scale e-commerce platform,
    we show that position debiasing can significantly improve assortment utilization,
    without any degradation in user engagement or financial metrics.
    This makes the ranking fairer and helps attract more partners or content providers,
    benefiting the customers and the business in the long term.
\end{abstract}

\ccsdesc[500]{Information systems~Recommender systems}

\keywords{Recommender Systems, Feedback Loop, Position Bias, Popularity}

\maketitle

\section{Introduction}

From a long-term strategy standpoint, a modern e-commerce platform aims to be a one-stop shop for all platform-related shopping needs.
That requires offering vast product selections, making it challenging for customers to find products aligned with their preferences and current needs.
To address this challenge, e-commerce platforms deploy personalized recommender and ranking systems
that nowadays play a central role in the customer shopping experience \cite{alamdari}.
Yet, there is an obstacle down this path: the effectiveness of those personalization systems is reduced by a naturally present feedback loop \cite{10.1145/3564284}.
As items ranked higher receive more user attention, the production recommender model creates a skew in the collected user interaction data in favor of itself.
This skew then impacts subsequent models as they are trained on the collected data, creating a repetitive cycle and reinforcing suboptimal model behaviors.
For example, it makes filter bubbles \cite{10.1145/3240323.3240372} and echo chambers in e-commerce \cite{echo-chamber} more persistent.

Note that the key driver behind the described feedback loop is the tendency of users to attend to some positions in the layout more than to others.
This phenomenon is referred to as \textit{position bias} \cite{10.1145/3564284, collins}. Position bias can lead to a lack of interaction with highly relevant items that are ranked low.

The literature on position bias largely focuses on improving the relevance of ranking and the associated theories
and experiments predict gains in user engagement when the respective methods are deployed \cite{10.1145/3018661.3018699, 10.1145/3366423.3380255}.
We argue however that depending on the strength of the position bias and the effectiveness of
the respective debiasing method, one may not observe such gains in the short term.
On one hand, the strength of the position bias might not be sufficient to impact relevance significantly.
On the other -- mitigating position bias is known to be difficult
as the associated methods often lack robustness to data sparsity \cite{10.1145/2911451.2911537},
exhibit high variance \cite{10.1145/3336191.3371783, pmlr-v48-schnabel16},
or suffer from interleaving biases \cite{10.1145/3340531.3412031}.

Yet, as this paper shows, debiasing has another potential benefit apart from driving engagement.
Namely, it can spread visibility and interactions more evenly across the assortment,
reducing the extra skew in the popularity of items incurred by the position bias.

Most relevant to our work are studies that have looked into how the feedback loop impacts popularity
\cite{10.1145/3340531.3412152, 10.1145/3447548.3467376, 10.1007/s11257-015-9165-3}.
In simulations, they show that the feedback loop tends to make already popular items even more popular
and less popular items even less popular, creating a ``rich get richer'' effect.
Effectively, it means that recorded sets of user interactions become more homogeneous \cite{10.1145/3306618.3314288, 10.1145/3240323.3240370}.

In this paper, we offer a theoretical view of how the feedback loop driven by the position bias affects item popularity.
We propose a model to quantify the skew in the popularity of items and the effect of the position bias on that skew.
Through offline and online experiments on our e-commerce platform,
we demonstrate that position debiasing can effectively spread visibility and interactions
more evenly across the assortment while leaving user engagement and financial metrics intact.

\section{Recommender system and popularity of items}\label{sec:popularity}

Consider a recommender system serving a stream of incoming user requests. In response to each of those requests,
the system displays a sequence of recommended items taken from a larger item vocabulary (assortment).
Presented with that sequence, the user observes some of its elements and interacts with those observed items that he or she finds relevant.
Over any fixed time frame, this process generates a number of interactions.
Naturally, different items in the vocabulary receive different number of interactions.
We use the term \emph{popularity} to refer to the share of interactions accumulated by a given item.

\begin{figure*}[ht]
    \centering
    \begin{subfigure}[t]{0.3\textwidth}
        \includegraphics[width=0.9\textwidth]{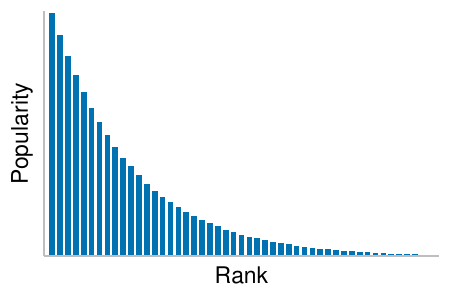}
        \caption{}
        \label{subfig:histogram}
    \end{subfigure}
    \begin{subfigure}[t]{0.3\textwidth}
        \includegraphics[width=0.9\textwidth]{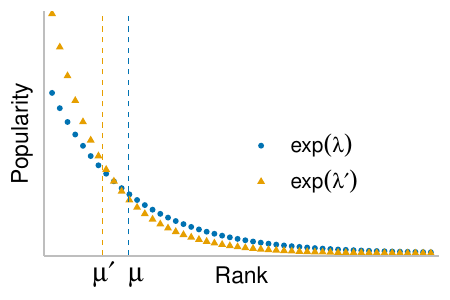}
        \caption{}
        \label{subfig:skew}
    \end{subfigure}
    \caption{
        Distribution of popularity of items.
        (a) Example of the histogram.
        (b) Discretized exponential distribution.
        The distribution with parameter $\lambda'$ has a stronger bias towards favoring more popular items
        illustrated by the expected values $\mu$ and $\mu'$ of the rank.
    }
    \Description{Example of the histogram to model the distribution of popularity of items.}
\end{figure*}

If we rank the items in the vocabulary by their popularity and then plot the popularity
as in Figure~\ref{subfig:histogram} (starting from the most popular item),
we will obtain a histogram that can be typically approximated with a long-tailed distribution.

In this and subsequent sections, we will consider the cases with and without the presence of the position bias
and argue about its effect on the skew of the popularity histogram.

First, consider a hypothetical case where position bias does not exist.
In this setting, the user interacts with a recommended item according to the probability of its relevance to the user.
For that, we denote by $D$ (without prime symbol) the interaction data
that represents the current iteration that maintains its natural bias in popularity
and \emph{remains unaffected by the feedback loop}.
This data is used by the recommender system to train and serve a model,
resulting in interactions that may be confounded by the exposure mechanism -- the feedback loop.
Consequently, data $D'$ is obtained in the subsequent iteration.

\subsection{Interaction model without position bias}

We follow the Plackett-Luce ranking model \cite{luce1977choice, plackett1975analysis} to describe the interaction model.
Consider a situation where we display an item $i$ given some presented ranking to the user $u$.
Let $p_{ui}$ denote the actual relevance of item $i$ to the user $u$
(i.e. the probability of the user interacting with an item provided the item has been observed).
To model user engagement with recommended items $R$ without position bias,
we interpret the selection of an item at time step $t$ from a presented ranking subset of items $\bm{r_t} \in R$
as sampling from the probability mass function $f_{I}(i \mpipe \bm{r_t})$ of a random variable $I_t$.
This random variable represents the selected item at time step $t$.
Formally, this is defined as follows:
\begin{equation}\label{eq:interaction-model-ubiased}
    \begin{aligned}
        f_{I}(i \mpipe \bm{r_t}) &= \frac{p_{ui}}{\sum_{j \in \bm{r_t}} p_{uj}}
        , \quad i \in  \bm{r_t},\\
        I_t &\sim f_{I}\left(i \mpipe \bm{r_t}\right).
    \end{aligned}
\end{equation}
These interactions $\left(I_t\right)$ together with the recommended items $R$ on the current iteration
will constitute data $D'$ on the next iteration.

We assume that the presented ranking, along with its interactions, can be viewed as a resampling process.
In the absence of feedback loop effects, these observations are expected to produce data with distributions
that are approximately equivalent, $D' \approx D$, neglecting randomness and user behavior/traffic changes.

This leads us to a key observation of how we model the effect of position bias on the popularity histogram,
which has remained unaffected by the feedback loop until now.
Specifically, we note that the collected interaction data can be viewed as a sample from a distribution over the vocabulary.

\section{Interaction model within the feedback loop}

\begin{figure}[ht]
    \centering
    \includegraphics[width=1.0\linewidth]{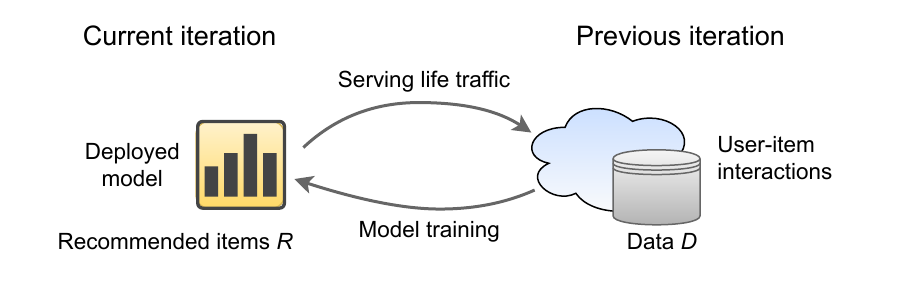}
    \caption{Model's life cycle within the feedback loop.}
    \Description{The model on the current iteration may induce a skew in the data on the next iteration.}
    \label{fig:iterations}
\end{figure}

A feedback loop is a mechanism that makes past ranking results appear in the data as more aligned with user preferences than they are.
It is called a loop because once a deployed model induces a skew in the data, it affects models trained on that data.
Those models get deployed, and the whole phenomenon repeats.
These iterations are illustrated in Figure~\ref{fig:iterations}.

The most prominent driver of the feedback loop is position bias.
Now, we present the case where the user behavior exhibits the position bias while interacting with the recommendation model,
which is trained on data $D$.
This means that the users may not observe all of the positions in the recommended list and may also pay different attention to different positions.
In the remainder of this section, we elaborate on the concept of position bias and define it formally.
Then, we extend the interaction model defined in (\ref{eq:interaction-model-ubiased}).

\subsection{Position bias}

Let $\one{C}$ be the indicator of the event that the user interacted with the item $i$ (e.g. clicked on it),
$\one{O}$ -- the indicator of the event that the user observed the displayed item $i$,
and $\mathrm{pos}(i)$ -- the position at which the item was displayed.
Define the relevance variable, $\one{V} \sim \mathrm{Bernoulli}\left(p_{ui}\right)$,
as the indicator of the event that the item is relevant to the user.
Under the standard \emph{examination hypothesis} \cite{craswell2008experimental},
a displayed item receives a click if and only if the user observes it and finds relevant, i.\,e.
\begin{equation}\label{eq:click-model}
    \one{C} = \one{O} \cdot \one{V}.
\end{equation}
From Eq.~\eqref{eq:click-model} it immediately follows that
\[
    \PR\left( \one{C} = 1 \mpipe \mathrm{pos}(i) = k \right) \leq \PR\left( \one{V} = 1 \mpipe \mathrm{pos}(i) = k \right).
\]
Then the position bias can be defined as the following ratio:
\[
    \mathrm{bias}(k) \coloneq {\PR\left( \one{C} = 1 \mpipe \mathrm{pos}(i) = k \right)} \mathbin{/} {\PR\left(\one{V} = 1 \mpipe \mathrm{pos}(i) = k \right)}.
\]
From Eq.~\eqref{eq:click-model}, it follows that
\[
    \PR\left( \one{C} = 1 \mpipe \mathrm{pos}(i) = k \right) = \PR\left(\one{C}\one{V} = 1 \mpipe \mathrm{pos}(i) = k \right),
\]
hence by the definition of conditional probability, the position bias is the probability of being clicked conditional on being relevant:
\[
    \mathrm{bias}(k) = \PR\left(\one{C} = 1 \mpipe \one{V} = 1, \mathrm{pos}(i) = k \right).
\]
Finally, assuming $O$ and $V$ are independent events conditional on the position, we have that
\begin{equation*}
    \PR\left( \one{C} = 1 \mpipe \mathrm{pos}(i) = k \right)
    = \PR\left( \one{O} = 1 \mpipe \mathrm{pos}(i) = k \right) \PR\left( \one{V} = 1 \mpipe \mathrm{pos}(i) = k \right),
\end{equation*}
and the bias is just the probability of observing the item $i$ displayed at the $k$-th position,
\[
    \mathrm{bias}(k) = \PR\left(\one{O} = 1 \mpipe \mathrm{pos}(i) = k \right).
\]
The position bias is commonly modeled as follows \cite{10.1145/3018661.3018699}:
\begin{equation}\label{eq:bias-joachims}
    \mathrm{bias}(k) \propto k^{-\beta},
\end{equation}
where the parameter $\beta$ controls the severity of bias.

\subsection{Interaction model}

To reason about the feedback loop dynamics of user behavior and algorithmic recommendations built
on observations $D$, we define a model of how users engage with recommended items $R$ considering the presence of position bias.
For that, we extend the model (\ref{eq:interaction-model-ubiased}) with the position bias defined earlier as follows:
\begin{equation}\label{eq:interaction-model-biased}
    \begin{aligned}
        f_{I}(i \mpipe \bm{r_t}) &= \frac{
            \mathrm{bias}(\mathrm{pos}(i \mpipe \bm{r_t})) \cdot p_{ui}
        }{
            \sum_{j \in \bm{r_t}} \mathrm{bias}(\mathrm{pos}(j \mpipe \bm{r_t})) \cdot p_{uj}
        }
        , \quad i \in  \bm{r_t},\\
        I_t &\sim f_{I}\left(i \mpipe \bm{r_t}\right),
    \end{aligned}
\end{equation}
where the item's selection is done by the user $u$ at time step $t$ given a presented ranking $\bm{r_t} \in R$.
This way, position bias introduces extra skew into sampling from the underlying distribution of the user judgment.
The interactions $\left(I_t\right)$ will produce biased data $D'$ with distributions different from $D$.

\section{Skew in the distribution of popularity of items}

In this section, we demonstrate and quantify the skew in the popularity histogram based on interaction data $D$.
As we explained earlier, we view such interaction data as a sample from a distribution over the vocabulary of items.
A common approach to model such a distribution is by discretizing a continuous density, like exponential distribution,
which gives a suitable approximation to a discrete empirical distribution.
We will use the assumption of an exponential distribution.

\subsection{Sample selection bias}

Let us first turn to the case where the position bias does not exist
and let $\lambda$ be the rate of the exponential distribution that approximates
the popularity histogram or, equivalently, the item sampling distribution in that case.

Define the random variable $X$ as the popularity rank of an item.
Now, the interaction model (\ref{eq:interaction-model-ubiased}) without position bias
results in sampling a random variable $X_t$ at time step $t$ from an exponential distribution with the rate parameter $\lambda$.
As a result, the current iteration produces observations $(X_t)$ along with interactions $(I_t)$.
These observations are expected to yield a similar distribution of popularity of items.

Next consider the interaction model (\ref{eq:interaction-model-biased}) with position bias.
The interpretation of the interaction data as an item sample applies in this scenario too but the sampling probabilities change.
We encounter biased sampling of $X_t$ because the sampling distribution is different from the target population $\exp\left(\lambda\right)$.
This is known as the sample selection bias \cite{10.1145/1015330.1015425}.
In this scenario, the probability density function of $X$ can be described as follows using the weighted exponential distribution:
\begin{equation}\label{eq:weighted-pdf}
    f_{X}(x) = \frac{
        w_{\mathrm{bias}}(x) \lambda e^{-\lambda x}
    }{
        \int_{0}^{\infty} w_{\mathrm{bias}}(x) \lambda e^{-\lambda x} \,dx
    }
    \propto \lambda'e^{-\lambda' x}
    , \quad x > 0,
\end{equation}
where $x$ corresponds to the popularity rank of an item and
$w_{\mathrm{bias}}(x)$
is a weighting function that describes the effect of the position bias on the popularity at different ranks.

Since the bias is typically a monotonically decreasing function of the position and popular items are generally shown in earlier positions,
it is natural to expect that $\lambda' > \lambda$ or, in other words, that the popularity histogram has a greater skew when the position bias is present.
Such positive skewness can be demonstrated by estimating a monotonic density under selection bias sampling \cite{Qin2017}.

\subsection{Influence on popularity}

Without correcting for the bias, the distribution of popularity of items will follow $\exp(\lambda')$ from Eq.~\eqref{eq:weighted-pdf},
exhibiting a stronger bias towards more popular items as depicted in Figure~\ref{subfig:skew}.
A commonly used method to correct for sample selection bias is the inverse probability-weighted method
originated from Horvitz and Thompson~\cite{bbbe4e97-7833-3faa-820c-4c61f82fe965},
akin to the inverse propensity scores method used in counterfactual learning-to-rank \cite{10.1145/3366423.3380255}.

In this manner, we show how the feedback loop driven by the position bias affects item popularity.
Addressing position bias mitigates its influence on popularity, effectively reducing the skew.
The validity of this popularity amplification is supported by simulations conducted in~\cite{10.1145/3340531.3412152},
where the sampling process, derived from user interactions, can be seen as a form of sample selection bias.
If we further assume that the deployment of a debiased model would make the popularity histogram look closer to the hypothetical case
in which the position bias is absent, we should expect the position debiasing to reduce the skew in the recorded popularity histogram.
This is exactly what we observed in our debiasing experiment that we present next.

At last, we quantify the skew in the popularity of items induced by the feedback loop
as the relative change
$\left(\lambda^{'} - \lambda\right) / {\lambda}$
between the actual parameter $\lambda$ and the biased parameter $\lambda^{'}$ of Eq.~\eqref{eq:weighted-pdf}.
The magnitude of the change reflects the strength of the feedback loop's influence on popularity.

\section{Experiments and results}

To address position bias and counteract the extra skew in the popularity of items influenced by the feedback loop,
we integrate position-aware learning \cite{10.1145/3041021.3054192} into the ranking model that powers both Browse and Search use cases of our
e-commerce platform's catalog.
This approach models positional information as a feature during training,
allowing the model to separate the impact of item position from its actual relevance.
Due to its simplicity, this method is widely used in practical applications \cite{10.1145/3298689.3347033}.
In contrast, another common approach -- using inverse propensity weighting transformations during training -- often results
in challenges related to the accuracy of the transformations \cite{10.1145/3041021.3054192} and high variance \cite{10.1145/3336191.3371783}.

\textbf{Dataset.} Our dataset consists of a sample of catalog sessions.
Each session consists of articles displayed in response to the customer's browse or search request.
This set of articles is combined with contextual data (e.g. market, device type, browsing category, etc),
user history prior to the time of the request (previous product clicks, add-to-cart and add-to-wishlist events, purchases),
and information about which of the displayed received an interaction from the customer.
The training dataset consists of 250 million sessions, involving 71 million unique customers across 25 markets,
with an average history length of 24 actions.
We split the sessions temporarily to create the training and test datasets to ensure no data leakage.

\begin{figure*}[ht]
    \centering
    \begin{subfigure}[t]{0.49\textwidth}
        \includegraphics[width=0.95\textwidth]{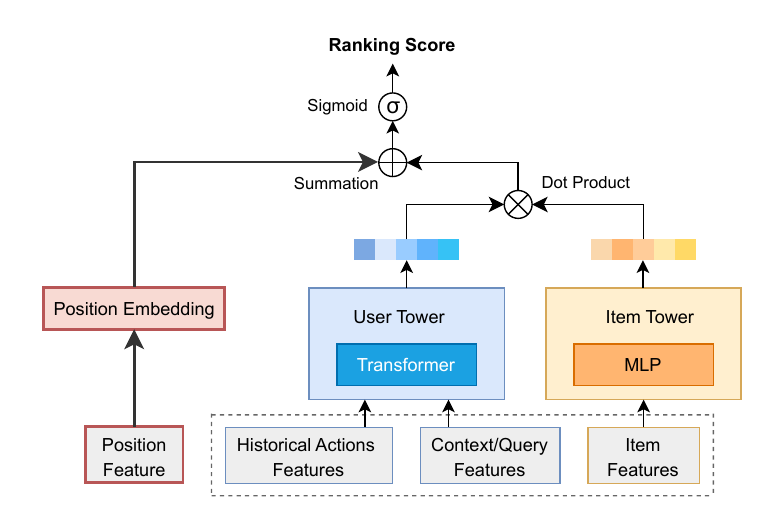}
        \caption{The model architecture with the position branch.}
        \label{subfig:debiasing-model}
    \end{subfigure}
    \begin{subfigure}[t]{0.49\textwidth}
        \includegraphics[width=0.8\textwidth]{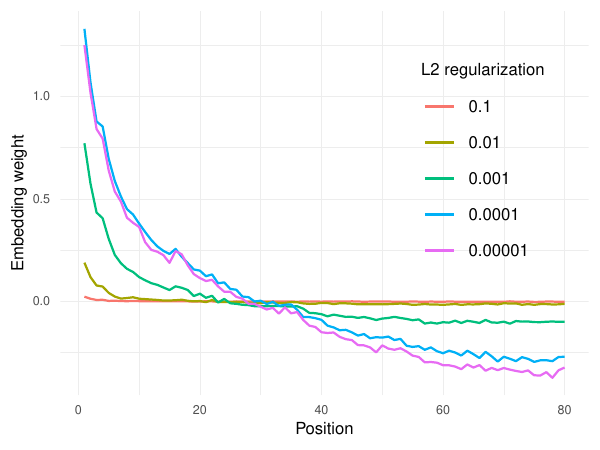}
        \caption{Embedding weight per position after training.}
        \label{subfig:weights}
    \end{subfigure}
    \caption{Position-aware learning.}
    \Description{The model architecture with a shallow position branch for debiasing.}
\end{figure*}

\textbf{Base Model.} We performed debiasing on top of an existing catalog ranking model.
This  model has a two-tower architecture \cite{youtuberec, twotowersampling}.
It did not have any mechanisms to remove position bias prior to our intervention.
The training objective of the model is to rank a given set of items.
This task is modeled as a pointwise prediction problem using binary cross entropy,
where we predict the probability of a customer performing a positive action.
Within this model, the user tower utilizes historical action sequences and contextual data to produce a user embedding,
while the item tower represents item embeddings.
Then, these embeddings are combined using a dot product operation to produce a score per item.
We use early stopping as a regularization technique to halt training when parameter updates no longer yield improvements on a validation set.
Figure~\ref{subfig:debiasing-model} depicts the final architecture, which includes the baseline model along with a newly added shallow position branch.

\textbf{Methodology.} Using the position-aware approach \cite{10.1145/3298689.3347033, 10.1145/3298689.3346997}, we add position information as a feature,
allowing the model to disentangle the influence of item position from the true relevance of the probability that a user would engage with an item.
During training, the model is conditioned on positions, while during serving, it becomes position-independent by setting a default value.
To prevent potential negative effects of correlation with other features,
we isolate this positional feature from the remaining features by using a shallow position branch.
Additionally, to prevent overfitting on position information, we apply L2 regularization to the position embeddings.
Figure~\ref{subfig:weights} illustrates the position embedding weights for each position based
on the regularization applied.
Higher weights indicate greater overfitting to position information during training.
Adding the position branch did not significantly change the training dynamics.
It slightly increased the loss, but the number of epochs required to reach saturation remained the same.

During the offline evaluation, we use inverse-propensity-weighted NDCG to measure relevance
and an average recommendation popularity metric to detect improvements in popularity skew.
In the online experiment, we also measure the effective catalog size.
The main metrics are described as follows:
\begin{itemize}
    \item Recall@k~\cite{10.1145/3460231.3478848}: Proportion of all relevant items within top-k items.
    \item Inverse Propensity Score weighted NDCG (IPS-NDCG@k) \cite{10.1145/3366423.3380255}:
    Assesses ranking effectiveness by considering the position of relevant items within the top-k list.
    All attributed items are considered as relevant and their relevance is weighted based on inverse propensity scores.
    The propensities were modeled as Eq.~\eqref{eq:bias-joachims}, and their severity $\beta$ was estimated
    using the Expectation-Maximization algorithm \cite{10.1145/3159652.3159732}.
    It is essentially a correction for bias in Eq.~\eqref{eq:weighted-pdf}.
    \item Average Recommendation Popularity within top k items (ARP@k) \cite{10.14778/2311906.2311916}:
    Measures the average popularity of recommended items in each list.
    For any item in the list, popularity is computed by the number of interactions
    accumulated in the preceding days.
    A higher ARP indicates a greater propensity for popular items within the recommendations.
    \item Effective Catalog Size (ECS@X) \cite{10.1145/2843948}:
    Measures the share of items that constitute X\% of all interactions aiming at
    describing the distribution of interactions.
\end{itemize}

\subsection{Offline experiments}

We offline evaluate the model on a holdout set containing instances (catalog sessions)
from the subsequent day, ensuring that the evaluation data remains unseen during the training phase.
Table~\ref{tab:grid-search} presents the results of a grid search for L2 regularization,
identifying the best model for further online experiments with an L2 value of 0.001.
After tuning the regularization, the offline evaluation showed no statistically significant improvement in IPS-NDCG (relevance),
but did demonstrate a notable improvement of -4.34\% in ARP,
with a statistically significant difference (p-value < 0.05).
Further adjustments in either direction did not improve the results.
Additionally, we included the performance of random and popularity-based baselines.
\begin{table}[htbp]
    \caption{Results of grid search for L2 regularization compared to the baseline model without position-aware learning. Best results for each metric are highlighted
    with an underscore.}
    \label{tab:grid-search}
    \centering
    \begin{tabular}{l|c|c|c}
        \toprule
        Model & Recall@6 & ARP@6 & IPS-NDCG@6 \\
        \midrule
        \textbf{L2 = 0.001} &  \textbf{\underline{-0.15\%}} & \textbf{-4.34\%} & \textbf{\underline{0.16\%}} \\
        L2 = 0.0001 &  -0.97\% & -8.66\% & -1.43\% \\
        L2 = 0.00001 &  -1.64\% & -14.68\% & -2.70\% \\
        \midrule
        Popularity &  -28.94\% & 129.10\% & -14.78\% \\
        Random &  -49.51\% & \underline{-48.19\%} & -27.11\% \\
        \bottomrule
    \end{tabular}
\end{table}

\subsection{Online experiment}

We conducted an online A/B test on the ranking use case, where we allocated equal traffic splits among variants
over several weeks to achieve the minimum detectable effect for the success KPI, with a p-value < 0.05.
The findings from offline experiments were consistent with the outcomes of the A/B test:
\begin{itemize}
    \item No statistically significant changes in the main KPIs, customer engagement and financial metrics,
    or either of the guardrail KPIs: net merchandise value after return per user and discovery return days per user.
    \item Decrease (i.e. improvement) in the popularity metric by 5.7\% in ARP@6.
    \item Increase (i.e. improvement) in the catalog items utilization by 3.1\% in ECS@10.
\end{itemize}
The results are further detailed in Table~\ref{tab:arp} and Table~\ref{tab:ecs}.
\begin{table}[htbp]
    \caption{The relative change of the average recommendation popularity (ARP@k).}
    \label{tab:arp}
    \centering
    \begin{tabular}{c|c|c|c|c}
        \toprule
        k & top-6 & top-12 & top-24 & top-84 \\
        \midrule
        ARP & -5.7\% &  -5.4\% & -4.9\% & -2.9\% \\
        \bottomrule
    \end{tabular}
\end{table}
\begin{table}[htbp]
    \caption{The relative change of the effective catalog size (ECS@X).}
    \label{tab:ecs}
    \centering
    \begin{tabular}{c|c|c|c|c|c|c|c|c|c}
        \toprule
        X & 0.1 & 0.2 & 0.3 & 0.4 & 0.5 & 0.6 & 0.7 & 0.8 & 0.9 \\
        \midrule
        ECS & 3.1\% & 2.2\% & 1.8\% & 1.4\% & 1.3\%& 1.1\%& 1.0\%& 0.8\%& 0.5\% \\
        \bottomrule
    \end{tabular}
\end{table}

The experiments demonstrate that position debiasing can effectively spread visibility and interactions more evenly across the assortment,
maintaining user engagement and financial metrics even when the strength of the position bias is not sufficient to impact relevance significantly.
This makes the ranking fairer and helps attract more partners or content providers, benefiting the customers and the business in the long term.

At last, following the deployment of the debiased model, we calculated the skew in the popularity of items as the relative change between
the distribution parameters before and after the model rollout, as defined in the previous section.
We approximated the distributions using exponential distribution with parameters,
determined through maximum likelihood distribution fitting.
The rollout of the debiased model resulted in a 2.5\% reduction in skew,
indicating a shift towards a more balanced distribution across items.
It's important to note that this comparison in skew is not rigorous due to inherent daily variations in factors
such as user traffic patterns, behavior distribution, introduction of new items, and other sources of randomness.
As a result, it serves as a supplementary metric to the main KPIs.

\section{Conclusion}

The paper has demonstrated that debiasing can spread visibility and interactions more evenly across the assortment
without hurting user engagement and financial metrics.
This makes the ranking fairer and helps attract more partners or content providers, benefiting the customers and the business in the long term.
We have provided a theoretical explanation of how the feedback loop, influenced by position bias, impacts popularity.
Through experiments on our e-commerce platform, we have showcased these findings.

\bibliographystyle{ACM-Reference-Format}
\bibliography{main}


\begin{thebibliography}{31}


\ifx \showCODEN    \undefined \def \showCODEN     #1{\unskip}     \fi
\ifx \showDOI      \undefined \def \showDOI       #1{#1}\fi
\ifx \showISBNx    \undefined \def \showISBNx     #1{\unskip}     \fi
\ifx \showISBNxiii \undefined \def \showISBNxiii  #1{\unskip}     \fi
\ifx \showISSN     \undefined \def \showISSN      #1{\unskip}     \fi
\ifx \showLCCN     \undefined \def \showLCCN      #1{\unskip}     \fi
\ifx \shownote     \undefined \def \shownote      #1{#1}          \fi
\ifx \showarticletitle \undefined \def \showarticletitle #1{#1}   \fi
\ifx \showURL      \undefined \def \showURL       {\relax}        \fi
\providecommand\bibfield[2]{#2}
\providecommand\bibinfo[2]{#2}
\providecommand\natexlab[1]{#1}
\providecommand\showeprint[2][]{arXiv:#2}

\bibitem[Alamdari et~al\mbox{.}(2020)]%
        {alamdari}
\bibfield{author}{\bibinfo{person}{Pegah~Malekpour Alamdari},
  \bibinfo{person}{Nima~Jafari Navimipour}, \bibinfo{person}{Mehdi
  Hosseinzadeh}, \bibinfo{person}{Ali~Asghar Safaei}, {and}
  \bibinfo{person}{Aso Darwesh}.} \bibinfo{year}{2020}\natexlab{}.
\newblock \showarticletitle{A Systematic Study on the Recommender Systems in
  the E-Commerce}.
\newblock \bibinfo{journal}{\emph{IEEE Access}}  \bibinfo{volume}{8}
  (\bibinfo{year}{2020}), \bibinfo{pages}{115694--115716}.
\newblock
\urldef\tempurl%
\url{https://doi.org/10.1109/ACCESS.2020.3002803}
\showDOI{\tempurl}


\bibitem[Chaney et~al\mbox{.}(2018)]%
        {10.1145/3240323.3240370}
\bibfield{author}{\bibinfo{person}{Allison J.~B. Chaney},
  \bibinfo{person}{Brandon~M. Stewart}, {and} \bibinfo{person}{Barbara~E.
  Engelhardt}.} \bibinfo{year}{2018}\natexlab{}.
\newblock \showarticletitle{How algorithmic confounding in recommendation
  systems increases homogeneity and decreases utility}. In
  \bibinfo{booktitle}{\emph{Proceedings of the 12th ACM Conference on
  Recommender Systems}} (Vancouver, British Columbia, Canada)
  \emph{(\bibinfo{series}{RecSys '18})}. \bibinfo{publisher}{Association for
  Computing Machinery}, \bibinfo{address}{New York, NY, USA},
  \bibinfo{pages}{224–232}.
\newblock
\showISBNx{9781450359016}
\urldef\tempurl%
\url{https://doi.org/10.1145/3240323.3240370}
\showDOI{\tempurl}


\bibitem[Chen et~al\mbox{.}(2023)]%
        {10.1145/3564284}
\bibfield{author}{\bibinfo{person}{Jiawei Chen}, \bibinfo{person}{Hande Dong},
  \bibinfo{person}{Xiang Wang}, \bibinfo{person}{Fuli Feng},
  \bibinfo{person}{Meng Wang}, {and} \bibinfo{person}{Xiangnan He}.}
  \bibinfo{year}{2023}\natexlab{}.
\newblock \showarticletitle{Bias and Debias in Recommender System: A Survey and
  Future Directions}.
\newblock \bibinfo{journal}{\emph{ACM Trans. Inf. Syst.}} \bibinfo{volume}{41},
  \bibinfo{number}{3}, Article \bibinfo{articleno}{67} (\bibinfo{date}{feb}
  \bibinfo{year}{2023}), \bibinfo{numpages}{39}~pages.
\newblock
\showISSN{1046-8188}
\urldef\tempurl%
\url{https://doi.org/10.1145/3564284}
\showDOI{\tempurl}


\bibitem[Collins et~al\mbox{.}(2018)]%
        {collins}
\bibfield{author}{\bibinfo{person}{Andrew Collins}, \bibinfo{person}{Dominika
  Tkaczyk}, \bibinfo{person}{Akiko Aizawa}, {and} \bibinfo{person}{Joeran
  Beel}.} \bibinfo{year}{2018}\natexlab{}.
\newblock \bibinfo{booktitle}{\emph{Position Bias in Recommender Systems for
  Digital Libraries}}.
\newblock \bibinfo{pages}{335--344}.
\newblock
\showISBNx{978-3-319-78104-4}
\urldef\tempurl%
\url{https://doi.org/10.1007/978-3-319-78105-1_37}
\showDOI{\tempurl}


\bibitem[Covington et~al\mbox{.}(2016)]%
        {youtuberec}
\bibfield{author}{\bibinfo{person}{Paul Covington}, \bibinfo{person}{Jay
  Adams}, {and} \bibinfo{person}{Emre Sargin}.}
  \bibinfo{year}{2016}\natexlab{}.
\newblock \showarticletitle{Deep Neural Networks for YouTube Recommendations}.
  In \bibinfo{booktitle}{\emph{Proceedings of the 10th ACM Conference on
  Recommender Systems}} (Boston, Massachusetts, USA)
  \emph{(\bibinfo{series}{RecSys '16})}. \bibinfo{publisher}{Association for
  Computing Machinery}, \bibinfo{address}{New York, NY, USA},
  \bibinfo{pages}{191–198}.
\newblock
\showISBNx{9781450340359}
\urldef\tempurl%
\url{https://doi.org/10.1145/2959100.2959190}
\showDOI{\tempurl}


\bibitem[Craswell et~al\mbox{.}(2008)]%
        {craswell2008experimental}
\bibfield{author}{\bibinfo{person}{Nick Craswell}, \bibinfo{person}{Onno
  Zoeter}, \bibinfo{person}{Michael Taylor}, {and} \bibinfo{person}{Bill
  Ramsey}.} \bibinfo{year}{2008}\natexlab{}.
\newblock \showarticletitle{An experimental comparison of click position-bias
  models}. In \bibinfo{booktitle}{\emph{Proceedings of the 2008 international
  conference on web search and data mining}}. \bibinfo{pages}{87--94}.
\newblock
\urldef\tempurl%
\url{https://doi.org/10.1145/1341531.1341545}
\showDOI{\tempurl}


\bibitem[Ge et~al\mbox{.}(2020)]%
        {echo-chamber}
\bibfield{author}{\bibinfo{person}{Yingqiang Ge}, \bibinfo{person}{Shuya Zhao},
  \bibinfo{person}{Honglu Zhou}, \bibinfo{person}{Changhua Pei},
  \bibinfo{person}{Fei Sun}, \bibinfo{person}{Wenwu Ou}, {and}
  \bibinfo{person}{Yongfeng Zhang}.} \bibinfo{year}{2020}\natexlab{}.
\newblock \showarticletitle{Understanding Echo Chambers in E-commerce
  Recommender Systems}. \bibinfo{pages}{2261--2270}.
\newblock
\urldef\tempurl%
\url{https://doi.org/10.1145/3397271.3401431}
\showDOI{\tempurl}


\bibitem[Gomez-Uribe and Hunt(2016)]%
        {10.1145/2843948}
\bibfield{author}{\bibinfo{person}{Carlos~A. Gomez-Uribe} {and}
  \bibinfo{person}{Neil Hunt}.} \bibinfo{year}{2016}\natexlab{}.
\newblock \showarticletitle{The Netflix Recommender System: Algorithms,
  Business Value, and Innovation}.
\newblock \bibinfo{journal}{\emph{ACM Trans. Manage. Inf. Syst.}}
  \bibinfo{volume}{6}, \bibinfo{number}{4}, Article \bibinfo{articleno}{13}
  (\bibinfo{date}{dec} \bibinfo{year}{2016}), \bibinfo{numpages}{19}~pages.
\newblock
\showISSN{2158-656X}
\urldef\tempurl%
\url{https://doi.org/10.1145/2843948}
\showDOI{\tempurl}


\bibitem[Guo et~al\mbox{.}(2019)]%
        {10.1145/3298689.3347033}
\bibfield{author}{\bibinfo{person}{Huifeng Guo}, \bibinfo{person}{Jinkai Yu},
  \bibinfo{person}{Qing Liu}, \bibinfo{person}{Ruiming Tang}, {and}
  \bibinfo{person}{Yuzhou Zhang}.} \bibinfo{year}{2019}\natexlab{}.
\newblock \showarticletitle{PAL: a position-bias aware learning framework for
  CTR prediction in live recommender systems}. In
  \bibinfo{booktitle}{\emph{Proceedings of the 13th ACM Conference on
  Recommender Systems}} (Copenhagen, Denmark) \emph{(\bibinfo{series}{RecSys
  '19})}. \bibinfo{publisher}{Association for Computing Machinery},
  \bibinfo{address}{New York, NY, USA}, \bibinfo{pages}{452–456}.
\newblock
\showISBNx{9781450362436}
\urldef\tempurl%
\url{https://doi.org/10.1145/3298689.3347033}
\showDOI{\tempurl}


\bibitem[Horvitz and Thompson(1952)]%
        {bbbe4e97-7833-3faa-820c-4c61f82fe965}
\bibfield{author}{\bibinfo{person}{D.~G. Horvitz} {and} \bibinfo{person}{D.~J.
  Thompson}.} \bibinfo{year}{1952}\natexlab{}.
\newblock \showarticletitle{A Generalization of Sampling Without Replacement
  From a Finite Universe}.
\newblock \bibinfo{journal}{\emph{J. Amer. Statist. Assoc.}}
  \bibinfo{volume}{47}, \bibinfo{number}{260} (\bibinfo{year}{1952}),
  \bibinfo{pages}{663--685}.
\newblock
\showISSN{01621459}
\urldef\tempurl%
\url{http://www.jstor.org/stable/2280784}
\showURL{%
\tempurl}


\bibitem[Jannach et~al\mbox{.}(2015)]%
        {10.1007/s11257-015-9165-3}
\bibfield{author}{\bibinfo{person}{Dietmar Jannach}, \bibinfo{person}{Lukas
  Lerche}, \bibinfo{person}{Iman Kamehkhosh}, {and} \bibinfo{person}{Michael
  Jugovac}.} \bibinfo{year}{2015}\natexlab{}.
\newblock \showarticletitle{What recommenders recommend: an analysis of
  recommendation biases and possible countermeasures}.
\newblock \bibinfo{journal}{\emph{User Modeling and User-Adapted Interaction}}
  \bibinfo{volume}{25}, \bibinfo{number}{5} (\bibinfo{date}{dec}
  \bibinfo{year}{2015}), \bibinfo{pages}{427–491}.
\newblock
\showISSN{0924-1868}
\urldef\tempurl%
\url{https://doi.org/10.1007/s11257-015-9165-3}
\showDOI{\tempurl}


\bibitem[Jiang et~al\mbox{.}(2019)]%
        {10.1145/3306618.3314288}
\bibfield{author}{\bibinfo{person}{Ray Jiang}, \bibinfo{person}{Silvia
  Chiappa}, \bibinfo{person}{Tor Lattimore}, \bibinfo{person}{Andr\'{a}s
  Gy\"{o}rgy}, {and} \bibinfo{person}{Pushmeet Kohli}.}
  \bibinfo{year}{2019}\natexlab{}.
\newblock \showarticletitle{Degenerate Feedback Loops in Recommender Systems}.
  In \bibinfo{booktitle}{\emph{Proceedings of the 2019 AAAI/ACM Conference on
  AI, Ethics, and Society}} (Honolulu, HI, USA) \emph{(\bibinfo{series}{AIES
  '19})}. \bibinfo{publisher}{Association for Computing Machinery},
  \bibinfo{address}{New York, NY, USA}, \bibinfo{pages}{383–390}.
\newblock
\showISBNx{9781450363242}
\urldef\tempurl%
\url{https://doi.org/10.1145/3306618.3314288}
\showDOI{\tempurl}


\bibitem[Joachims et~al\mbox{.}(2017)]%
        {10.1145/3018661.3018699}
\bibfield{author}{\bibinfo{person}{Thorsten Joachims}, \bibinfo{person}{Adith
  Swaminathan}, {and} \bibinfo{person}{Tobias Schnabel}.}
  \bibinfo{year}{2017}\natexlab{}.
\newblock \showarticletitle{Unbiased Learning-to-Rank with Biased Feedback}. In
  \bibinfo{booktitle}{\emph{Proceedings of the Tenth ACM International
  Conference on Web Search and Data Mining}} (Cambridge, United Kingdom)
  \emph{(\bibinfo{series}{WSDM '17})}. \bibinfo{publisher}{Association for
  Computing Machinery}, \bibinfo{address}{New York, NY, USA},
  \bibinfo{pages}{781–789}.
\newblock
\showISBNx{9781450346757}
\urldef\tempurl%
\url{https://doi.org/10.1145/3018661.3018699}
\showDOI{\tempurl}


\bibitem[Ling et~al\mbox{.}(2017)]%
        {10.1145/3041021.3054192}
\bibfield{author}{\bibinfo{person}{Xiaoliang Ling}, \bibinfo{person}{Weiwei
  Deng}, \bibinfo{person}{Chen Gu}, \bibinfo{person}{Hucheng Zhou},
  \bibinfo{person}{Cui Li}, {and} \bibinfo{person}{Feng Sun}.}
  \bibinfo{year}{2017}\natexlab{}.
\newblock \showarticletitle{Model Ensemble for Click Prediction in Bing Search
  Ads}. In \bibinfo{booktitle}{\emph{Proceedings of the 26th International
  Conference on World Wide Web Companion}} (Perth, Australia)
  \emph{(\bibinfo{series}{WWW '17 Companion})}.
  \bibinfo{publisher}{International World Wide Web Conferences Steering
  Committee}, \bibinfo{address}{Republic and Canton of Geneva, CHE},
  \bibinfo{pages}{689–698}.
\newblock
\showISBNx{9781450349147}
\urldef\tempurl%
\url{https://doi.org/10.1145/3041021.3054192}
\showDOI{\tempurl}


\bibitem[Luce(1977)]%
        {luce1977choice}
\bibfield{author}{\bibinfo{person}{R~Duncan Luce}.}
  \bibinfo{year}{1977}\natexlab{}.
\newblock \showarticletitle{The choice axiom after twenty years}.
\newblock \bibinfo{journal}{\emph{Journal of mathematical psychology}}
  \bibinfo{volume}{15}, \bibinfo{number}{3} (\bibinfo{year}{1977}),
  \bibinfo{pages}{215--233}.
\newblock
\urldef\tempurl%
\url{https://doi.org/10.1016/0022-2496%2877%2990032-3}
\showDOI{\tempurl}


\bibitem[Mansoury et~al\mbox{.}(2020)]%
        {10.1145/3340531.3412152}
\bibfield{author}{\bibinfo{person}{Masoud Mansoury}, \bibinfo{person}{Himan
  Abdollahpouri}, \bibinfo{person}{Mykola Pechenizkiy},
  \bibinfo{person}{Bamshad Mobasher}, {and} \bibinfo{person}{Robin Burke}.}
  \bibinfo{year}{2020}\natexlab{}.
\newblock \showarticletitle{Feedback Loop and Bias Amplification in Recommender
  Systems}. In \bibinfo{booktitle}{\emph{Proceedings of the 29th ACM
  International Conference on Information \& Knowledge Management}} (Virtual
  Event, Ireland) \emph{(\bibinfo{series}{CIKM '20})}.
  \bibinfo{publisher}{Association for Computing Machinery},
  \bibinfo{address}{New York, NY, USA}, \bibinfo{pages}{2145–2148}.
\newblock
\showISBNx{9781450368599}
\urldef\tempurl%
\url{https://doi.org/10.1145/3340531.3412152}
\showDOI{\tempurl}


\bibitem[Ovaisi et~al\mbox{.}(2020)]%
        {10.1145/3366423.3380255}
\bibfield{author}{\bibinfo{person}{Zohreh Ovaisi}, \bibinfo{person}{Ragib
  Ahsan}, \bibinfo{person}{Yifan Zhang}, \bibinfo{person}{Kathryn Vasilaky},
  {and} \bibinfo{person}{Elena Zheleva}.} \bibinfo{year}{2020}\natexlab{}.
\newblock \showarticletitle{Correcting for Selection Bias in Learning-to-rank
  Systems}. In \bibinfo{booktitle}{\emph{Proceedings of The Web Conference
  2020}} (Taipei, Taiwan) \emph{(\bibinfo{series}{WWW '20})}.
  \bibinfo{publisher}{Association for Computing Machinery},
  \bibinfo{address}{New York, NY, USA}, \bibinfo{pages}{1863–1873}.
\newblock
\showISBNx{9781450370233}
\urldef\tempurl%
\url{https://doi.org/10.1145/3366423.3380255}
\showDOI{\tempurl}


\bibitem[Plackett(1975)]%
        {plackett1975analysis}
\bibfield{author}{\bibinfo{person}{Robin~L Plackett}.}
  \bibinfo{year}{1975}\natexlab{}.
\newblock \showarticletitle{The analysis of permutations}.
\newblock \bibinfo{journal}{\emph{Journal of the Royal Statistical Society
  Series C: Applied Statistics}} \bibinfo{volume}{24}, \bibinfo{number}{2}
  (\bibinfo{year}{1975}), \bibinfo{pages}{193--202}.
\newblock
\urldef\tempurl%
\url{https://doi.org/10.2307/2346567}
\showDOI{\tempurl}


\bibitem[Qin(2017)]%
        {Qin2017}
\bibfield{author}{\bibinfo{person}{Jing Qin}.} \bibinfo{year}{2017}\natexlab{}.
\newblock \bibinfo{booktitle}{\emph{Length Biased Sampling, Multiplicative
  Censoring and Survival Analysis}}.
\newblock \bibinfo{publisher}{Springer Singapore},
  \bibinfo{address}{Singapore}, \bibinfo{pages}{519--558}.
\newblock
\showISBNx{978-981-10-4856-2}
\urldef\tempurl%
\url{https://doi.org/10.1007/978-981-10-4856-2_25}
\showDOI{\tempurl}


\bibitem[Saito et~al\mbox{.}(2020)]%
        {10.1145/3336191.3371783}
\bibfield{author}{\bibinfo{person}{Yuta Saito}, \bibinfo{person}{Suguru
  Yaginuma}, \bibinfo{person}{Yuta Nishino}, \bibinfo{person}{Hayato Sakata},
  {and} \bibinfo{person}{Kazuhide Nakata}.} \bibinfo{year}{2020}\natexlab{}.
\newblock \showarticletitle{Unbiased Recommender Learning from
  Missing-Not-At-Random Implicit Feedback}. In
  \bibinfo{booktitle}{\emph{Proceedings of the 13th International Conference on
  Web Search and Data Mining}} (Houston, TX, USA) \emph{(\bibinfo{series}{WSDM
  '20})}. \bibinfo{publisher}{Association for Computing Machinery},
  \bibinfo{address}{New York, NY, USA}, \bibinfo{pages}{501–509}.
\newblock
\showISBNx{9781450368223}
\urldef\tempurl%
\url{https://doi.org/10.1145/3336191.3371783}
\showDOI{\tempurl}


\bibitem[Schnabel et~al\mbox{.}(2016)]%
        {pmlr-v48-schnabel16}
\bibfield{author}{\bibinfo{person}{Tobias Schnabel}, \bibinfo{person}{Adith
  Swaminathan}, \bibinfo{person}{Ashudeep Singh}, \bibinfo{person}{Navin
  Chandak}, {and} \bibinfo{person}{Thorsten Joachims}.}
  \bibinfo{year}{2016}\natexlab{}.
\newblock \showarticletitle{Recommendations as Treatments: Debiasing Learning
  and Evaluation}. In \bibinfo{booktitle}{\emph{Proceedings of The 33rd
  International Conference on Machine Learning}}
  \emph{(\bibinfo{series}{Proceedings of Machine Learning Research},
  Vol.~\bibinfo{volume}{48})}, \bibfield{editor}{\bibinfo{person}{Maria~Florina
  Balcan} {and} \bibinfo{person}{Kilian~Q. Weinberger}} (Eds.).
  \bibinfo{publisher}{PMLR}, \bibinfo{address}{New York, New York, USA},
  \bibinfo{pages}{1670--1679}.
\newblock
\urldef\tempurl%
\url{https://proceedings.mlr.press/v48/schnabel16.html}
\showURL{%
\tempurl}


\bibitem[Steck(2018)]%
        {10.1145/3240323.3240372}
\bibfield{author}{\bibinfo{person}{Harald Steck}.}
  \bibinfo{year}{2018}\natexlab{}.
\newblock \showarticletitle{Calibrated recommendations}. In
  \bibinfo{booktitle}{\emph{Proceedings of the 12th ACM Conference on
  Recommender Systems}} (Vancouver, British Columbia, Canada)
  \emph{(\bibinfo{series}{RecSys '18})}. \bibinfo{publisher}{Association for
  Computing Machinery}, \bibinfo{address}{New York, NY, USA},
  \bibinfo{pages}{154–162}.
\newblock
\showISBNx{9781450359016}
\urldef\tempurl%
\url{https://doi.org/10.1145/3240323.3240372}
\showDOI{\tempurl}


\bibitem[Tamm et~al\mbox{.}(2021)]%
        {10.1145/3460231.3478848}
\bibfield{author}{\bibinfo{person}{Yan-Martin Tamm}, \bibinfo{person}{Rinchin
  Damdinov}, {and} \bibinfo{person}{Alexey Vasilev}.}
  \bibinfo{year}{2021}\natexlab{}.
\newblock \showarticletitle{Quality Metrics in Recommender Systems: Do We
  Calculate Metrics Consistently?}. In \bibinfo{booktitle}{\emph{Proceedings of
  the 15th ACM Conference on Recommender Systems}} (Amsterdam, Netherlands)
  \emph{(\bibinfo{series}{RecSys '21})}. \bibinfo{publisher}{Association for
  Computing Machinery}, \bibinfo{address}{New York, NY, USA},
  \bibinfo{pages}{708–713}.
\newblock
\showISBNx{9781450384582}
\urldef\tempurl%
\url{https://doi.org/10.1145/3460231.3478848}
\showDOI{\tempurl}


\bibitem[Vardasbi et~al\mbox{.}(2020)]%
        {10.1145/3340531.3412031}
\bibfield{author}{\bibinfo{person}{Ali Vardasbi}, \bibinfo{person}{Harrie
  Oosterhuis}, {and} \bibinfo{person}{Maarten de Rijke}.}
  \bibinfo{year}{2020}\natexlab{}.
\newblock \showarticletitle{When Inverse Propensity Scoring does not Work:
  Affine Corrections for Unbiased Learning to Rank}. In
  \bibinfo{booktitle}{\emph{Proceedings of the 29th ACM International
  Conference on Information \& Knowledge Management}} (Virtual Event, Ireland)
  \emph{(\bibinfo{series}{CIKM '20})}. \bibinfo{publisher}{Association for
  Computing Machinery}, \bibinfo{address}{New York, NY, USA},
  \bibinfo{pages}{1475–1484}.
\newblock
\showISBNx{9781450368599}
\urldef\tempurl%
\url{https://doi.org/10.1145/3340531.3412031}
\showDOI{\tempurl}


\bibitem[Wang et~al\mbox{.}(2016)]%
        {10.1145/2911451.2911537}
\bibfield{author}{\bibinfo{person}{Xuanhui Wang}, \bibinfo{person}{Michael
  Bendersky}, \bibinfo{person}{Donald Metzler}, {and} \bibinfo{person}{Marc
  Najork}.} \bibinfo{year}{2016}\natexlab{}.
\newblock \showarticletitle{Learning to Rank with Selection Bias in Personal
  Search}. In \bibinfo{booktitle}{\emph{Proceedings of the 39th International
  ACM SIGIR Conference on Research and Development in Information Retrieval}}
  (Pisa, Italy) \emph{(\bibinfo{series}{SIGIR '16})}.
  \bibinfo{publisher}{Association for Computing Machinery},
  \bibinfo{address}{New York, NY, USA}, \bibinfo{pages}{115–124}.
\newblock
\showISBNx{9781450340694}
\urldef\tempurl%
\url{https://doi.org/10.1145/2911451.2911537}
\showDOI{\tempurl}


\bibitem[Wang et~al\mbox{.}(2018)]%
        {10.1145/3159652.3159732}
\bibfield{author}{\bibinfo{person}{Xuanhui Wang}, \bibinfo{person}{Nadav
  Golbandi}, \bibinfo{person}{Michael Bendersky}, \bibinfo{person}{Donald
  Metzler}, {and} \bibinfo{person}{Marc Najork}.}
  \bibinfo{year}{2018}\natexlab{}.
\newblock \showarticletitle{Position Bias Estimation for Unbiased Learning to
  Rank in Personal Search}. In \bibinfo{booktitle}{\emph{Proceedings of the
  Eleventh ACM International Conference on Web Search and Data Mining}} (Marina
  Del Rey, CA, USA) \emph{(\bibinfo{series}{WSDM '18})}.
  \bibinfo{publisher}{Association for Computing Machinery},
  \bibinfo{address}{New York, NY, USA}, \bibinfo{pages}{610–618}.
\newblock
\showISBNx{9781450355810}
\urldef\tempurl%
\url{https://doi.org/10.1145/3159652.3159732}
\showDOI{\tempurl}


\bibitem[Yang et~al\mbox{.}(2020)]%
        {twotowersampling}
\bibfield{author}{\bibinfo{person}{Ji Yang}, \bibinfo{person}{Xinyang Yi},
  \bibinfo{person}{Derek Zhiyuan~Cheng}, \bibinfo{person}{Lichan Hong},
  \bibinfo{person}{Yang Li}, \bibinfo{person}{Simon Xiaoming~Wang},
  \bibinfo{person}{Taibai Xu}, {and} \bibinfo{person}{Ed~H. Chi}.}
  \bibinfo{year}{2020}\natexlab{}.
\newblock \showarticletitle{Mixed Negative Sampling for Learning Two-tower
  Neural Networks in Recommendations}. In \bibinfo{booktitle}{\emph{Companion
  Proceedings of the Web Conference 2020}} (Taipei, Taiwan)
  \emph{(\bibinfo{series}{WWW '20})}. \bibinfo{publisher}{Association for
  Computing Machinery}, \bibinfo{address}{New York, NY, USA},
  \bibinfo{pages}{441–447}.
\newblock
\showISBNx{9781450370240}
\urldef\tempurl%
\url{https://doi.org/10.1145/3366424.3386195}
\showDOI{\tempurl}


\bibitem[Yin et~al\mbox{.}(2012)]%
        {10.14778/2311906.2311916}
\bibfield{author}{\bibinfo{person}{Hongzhi Yin}, \bibinfo{person}{Bin Cui},
  \bibinfo{person}{Jing Li}, \bibinfo{person}{Junjie Yao}, {and}
  \bibinfo{person}{Chen Chen}.} \bibinfo{year}{2012}\natexlab{}.
\newblock \showarticletitle{Challenging the long tail recommendation}.
\newblock \bibinfo{journal}{\emph{Proc. VLDB Endow.}} \bibinfo{volume}{5},
  \bibinfo{number}{9} (\bibinfo{date}{may} \bibinfo{year}{2012}),
  \bibinfo{pages}{896–907}.
\newblock
\showISSN{2150-8097}
\urldef\tempurl%
\url{https://doi.org/10.14778/2311906.2311916}
\showDOI{\tempurl}


\bibitem[Zadrozny(2004)]%
        {10.1145/1015330.1015425}
\bibfield{author}{\bibinfo{person}{Bianca Zadrozny}.}
  \bibinfo{year}{2004}\natexlab{}.
\newblock \showarticletitle{Learning and evaluating classifiers under sample
  selection bias}. In \bibinfo{booktitle}{\emph{Proceedings of the Twenty-First
  International Conference on Machine Learning}} (Banff, Alberta, Canada)
  \emph{(\bibinfo{series}{ICML '04})}. \bibinfo{publisher}{Association for
  Computing Machinery}, \bibinfo{address}{New York, NY, USA},
  \bibinfo{pages}{114}.
\newblock
\showISBNx{1581138385}
\urldef\tempurl%
\url{https://doi.org/10.1145/1015330.1015425}
\showDOI{\tempurl}


\bibitem[Zhao et~al\mbox{.}(2019)]%
        {10.1145/3298689.3346997}
\bibfield{author}{\bibinfo{person}{Zhe Zhao}, \bibinfo{person}{Lichan Hong},
  \bibinfo{person}{Li Wei}, \bibinfo{person}{Jilin Chen},
  \bibinfo{person}{Aniruddh Nath}, \bibinfo{person}{Shawn Andrews},
  \bibinfo{person}{Aditee Kumthekar}, \bibinfo{person}{Maheswaran
  Sathiamoorthy}, \bibinfo{person}{Xinyang Yi}, {and} \bibinfo{person}{Ed
  Chi}.} \bibinfo{year}{2019}\natexlab{}.
\newblock \showarticletitle{Recommending what video to watch next: a multitask
  ranking system}. In \bibinfo{booktitle}{\emph{Proceedings of the 13th ACM
  Conference on Recommender Systems}} (Copenhagen, Denmark)
  \emph{(\bibinfo{series}{RecSys '19})}. \bibinfo{publisher}{Association for
  Computing Machinery}, \bibinfo{address}{New York, NY, USA},
  \bibinfo{pages}{43–51}.
\newblock
\showISBNx{9781450362436}
\urldef\tempurl%
\url{https://doi.org/10.1145/3298689.3346997}
\showDOI{\tempurl}


\bibitem[Zhu et~al\mbox{.}(2021)]%
        {10.1145/3447548.3467376}
\bibfield{author}{\bibinfo{person}{Ziwei Zhu}, \bibinfo{person}{Yun He},
  \bibinfo{person}{Xing Zhao}, {and} \bibinfo{person}{James Caverlee}.}
  \bibinfo{year}{2021}\natexlab{}.
\newblock \showarticletitle{Popularity Bias in Dynamic Recommendation}. In
  \bibinfo{booktitle}{\emph{Proceedings of the 27th ACM SIGKDD Conference on
  Knowledge Discovery \& Data Mining}} (Virtual Event, Singapore)
  \emph{(\bibinfo{series}{KDD '21})}. \bibinfo{publisher}{Association for
  Computing Machinery}, \bibinfo{address}{New York, NY, USA},
  \bibinfo{pages}{2439–2449}.
\newblock
\showISBNx{9781450383325}
\urldef\tempurl%
\url{https://doi.org/10.1145/3447548.3467376}
\showDOI{\tempurl}


\end{thebibliography}

\end{document}